\documentclass[twocolumn,superscriptaddress,notitlepage, 11pt,longbibliography,reprint]{revtex4-1}
\usepackage{amsmath,amssymb,graphicx}

\begin{document}

\title{Refractive index profile tailoring of multimode optical fibers for the spatial and spectral shaping of parametric sidebands}

\author{Katarzyna Krupa}
\affiliation{Dipartimento di Ingegneria dell'Informazione, Universit\`a di Brescia, via Branze 38, 25123, Brescia, Italy}

\author{Vincent Couderc}
\affiliation{Universit\'e de Limoges, XLIM, UMR CNRS 7252, 123 Av. A. Thomas, 87060 Limoges, France}
  
\author{Alessandro Tonello}
\affiliation{Universit\'e de Limoges, XLIM, UMR CNRS 7252, 123 Av. A. Thomas, 87060 Limoges, France}

\author{Daniele Modotto}
\affiliation{Dipartimento di Ingegneria dell'Informazione, Universit\`a di Brescia, via Branze 38, 25123, Brescia, Italy}

\author{Alain Barth\'el\'emy}
\affiliation{Universit\'e de Limoges, XLIM, UMR CNRS 7252, 123 Av. A. Thomas, 87060 Limoges, France}

\author{Guy Millot}
\affiliation{Universit\'e Bourgogne Franche-Comt\'e, ICB, UMR CNRS 6303, 9 Av. A. Savary, 21078 Dijon, France}

\author{Stefan Wabnitz}
\affiliation{DIET, Sapienza Universit\`a di Roma, via Eudossiana 18, 00184 Rome, Italy}
\affiliation{Dipartimento di Ingegneria dell'Informazione, Universit\`a di Brescia, via Branze 38, 25123, Brescia, Italy}
\affiliation{Istituto Nazionale di Ottica del Consiglio Nazionale delle Ricerche (INO-CNR), Via Branze 45, 25123 Brescia, Italy}

\email[]{Corresponding author: katarzyna.krupa@unibs.it} 

\begin{abstract}
We introduce the concept of spatial and spectral control of nonlinear parametric sidebands in multimode optical fibers by tailoring their linear refractive index profile. In all cases, the pump experiences Kerr self-cleaning, leading to a bell-shaped profile close to the fundamental mode. Geometric parametric instability, owing to quasi-phase-matching from the grating generated via the Kerr effect by pump self-imaging, leads to frequency multicasting of beam self-cleaning across a wideband array of sidebands. Our experiments show that introducing a gaussian dip in the refractive index profile of a graded index fiber permits to dramatically change the spatial content of spectral sidebands into higher-order modes. This is due to the breaking of oscillation synchronism among fundamental and higher-order modes. Hence modal-four-wave mixing prevails over geometric parametric instability as the main sideband generation mechanism. Observations agree well with theoretical predictions based on a perturbative analysis, and with full numerical solutions of the $(3D+1)$ nonlinear Schr\"odinger equation. 
\end{abstract}

\maketitle

\section{Introduction}
\label{section_one}
The last few years witnessed a strong renewed interest in the research on controllable spatio-temporal beam shaping with multimode optical fibers. In particular, graded-index multimode fibers (GRIN MMFs), thanks to their reduced modal dispersion which permits relatively long interaction lengths among the guided modes, were found to host a rich variety of spatiotemporal dynamical effects \cite{Wright2015R31,Picozzi2015R30,bib:SPIEProc}. Among these, we may cite multimode solitons \cite{Renninger2012R31}, ultra-wideband spectral sideband series generation from either oscillating multimode solitons \cite{Wright2015R31,Wright2015R29} or quasi continuous-wave (CW) pulses (also known as geometric parametric instability, GPI) \cite{Longhi2003R27,bib:KrupaPRLGPI,WrightNP2016,KrupaPRA}, intermodal four-wave mixing (IMFWM) \cite{bib:MafiJosab:16,bib:Dupiol:17}, beam self-cleaning in passive \cite{bib:KrupaPRLGPI,bib:KrupaNatPhot,LiuKerr,WrightNP2016} and active MMFs \cite{GuenardOpex}, supercontinuum generation \cite{Galmiche:16,bib:SCKrupa}, spatio-temporal mode-locking \cite{GuenardOpex2,WiseScience}, and wavefront shaping control of frequency conversion processes \cite{bib:Tzang:2018}, to name a few.  

It remains yet to explore the possibility to control spatio-temporal beam shaping in MMFs by a suitable tailoring of the refractive index profile of the fiber. 
In this paper we address this issue, and unveil the central 
role played by the refractive index profile for the control of spatial and spectral properties of nonlinear wave propagation in MMFs. 
In particular, we reveal that a relatively small localized depression (or dip) in the refractive index of a GRIN fiber has a significant impact both on Kerr beam self-cleaning, and most strikingly, on parametric sideband generation.
 
It is well known that any deviation of the refractive index profile of a GRIN MMF from the ideal parabolic shape has direct consequences on the spacing among propagation constants of its modes \cite{bib:Meunier:1980,bib:Sammut:1978,bib:Khular:1977}. In this work, we show that this simple consideration has dramatic consequences on the mode selection properties of parametric sidebands, as a result of the disruption of the collective self-imaging effect, which is a typical characteristic of multimode wave propagation in ideal GRIN fibers.

In order to generate a parabolic index profile, optical fibers are drawn from preforms, which are often obtained by chemical vapour deposition of dopants. A central index dip may appear in the deposition process, owing to the volatilization of silica, and dopants in the innermost layers \cite{bib:Khular:1977}. This effect is usually unwanted, and it can be circumvented by increasing the concentration of the more volatile dopant. 

Interestingly, the linear propagation equation describing light propagation in  GRIN MMF in the presence of diffraction and with a parabolic refractive index profile is fully analogous to the Schr\"odinger equation governing the state of a quantum system under the influence of a harmonic potential. A longitudinally invariant modification of the refractive index is then analogous to a time-invariant perturbation of the harmonic oscillator. From this viewpoint, the presence of a 
local depression of the refractive index or dip is equivalent to perturbing the harmonic potential with a barrier. This analogy permits a direct link with other physical problems which have been often studied in the past. By fitting the barrier with a gaussian function, it was possible to study the inversion of ammonia \cite{bib:Swalen:1962}. Notably, the harmonic potential with a gaussian barrier has been considered since a long time in the context of Bose-Einstein condensates (BEC). Solitons were also studied in the context of harmonic potentials with a gaussian barrier \cite{bib:Cuevas:2013}.

The introduction of a perturbation to the harmonic quantum potential leads to a splitting of energy levels (see for instance \cite{bib:Atkins}). In the case of classical optical waves in a fiber, where the role of time is replaced by the fiber length and propagation constants play the role of energy levels, the corresponding shift of propagation constants has direct consequences on phase-matching of four-wave mixing, hence on spatial and spectral properties of waves scattered through parametric processes. 
In the context of multimode nonlinear parametric sideband generation, it was recently pointed out 
that an imperfect parabolic shape could lead to splitting of propagation constants within nearly degenerate mode groups \cite{bib:Tzang:2018sup}. Moreover, the presence of stable soliton families was theoretically predicted, when the graded index profile is modified by the Kerr effect induced by light itself  \cite{Kivshar2018multimode}.
 
In this work we will show that a central index dip in the refractive index profile of GRIN fibers, a feature that one would typically consider as a defect, may in fact be exploited to inspire new ways of shaping the spatial and spectral domains of light waves generated by four-wave mixing. In GRIN fibers with an ideal parabolic index profile, GPI is pumped by a bell-shaped self-cleaned beam, which translates its spatial profile into all sidebands. These sidebands are generated by quasi-phase-matching (QPM) via the dynamic grating resulting from Kerr effect and pump self-imaging. The presence of a refractive index dip  breaks the oscillation synchronism among fundamental and higher-order modes, which spoils QPM between the pump and sidebands. As a result, to obtain phase-matching the sidebands and the pump need to be carried by different modes of the fiber. In the presence of a refractive index dip, IMFWM prevails over GPI as the sideband generation mechanism. Quite remarkably, the sideband positions with either GPI or IMFWM are exactly the same. 

\begin{figure}[hbt!]
\centering\includegraphics[width=7cm]{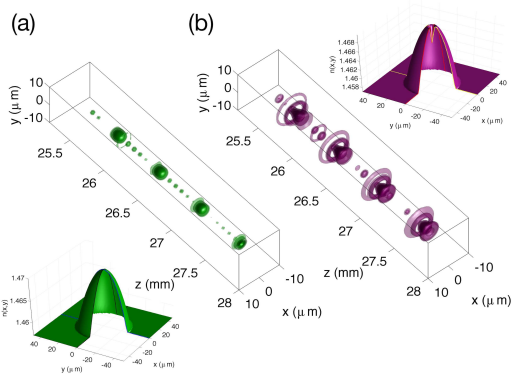}
\caption{Iso-intensity surfaces identifying the local variation of the refractive index induced by the Kerr effect, and obtained for $\delta n=0$ (a) and  $\delta n=-0.004$ (b). \label{fig5}}
\end{figure}

\section{Theoretical analysis}
\label{section_two}
In an ideal GRIN fiber, the regular spacing of modal propagation constants is at the origin of the periodic, with a sub-millimeter period, contraction of the beam waist ({\it i.e.}, the self-imaging effect) and the consequent periodic increase of the local intensity level. Since the refractive index is intensity dependent, the resulting long-period light-induced index grating may in turn generate an efficient quasi-phase matching mechanism for sideband waves. Wright {\it et al.} \cite{Wright2015R31,Wright2015R29}, for instance, observed in the anomalous dispersion regime the efficient coupling of multimode fiber solitons with multiple resonant dispersive waves, thanks to the spatiotemporal periodic oscillations of soliton beams. Whereas in the normal dispersion regime, Krupa et al. \cite{bib:KrupaPRLGPI} observed the break-up of  a nanosecond pulse leading to the generation of multiple sidebands carried by bell-shaped beams, as a result of GPI, that was previously predicted by S. Longhi \cite{Longhi2003R27}.
It is important to consider the nonlinear spatial beam reshaping, which accompanies GPI-induced sideband generation. GPI occurs at power values larger than the threshold for self-cleaning of the pump beam. Hence the pump beam is mostly carried by a bell-shaped transverse profile, close to the fundamental mode of the GRIN MMF\cite{bib:KrupaPRLGPI,bib:KrupaNatPhot}. Since the generation of the various sidebands is a result of QPM with the different orders of the light-induced index grating, there is no additional need for the sidebands to be carried by higher-order modes for a phase-matching of the FWM process. Spatial multicasting of the pump beam occurs: all sidebands are generated with the same bell-shaped transverse beam profile as the self-cleaned pump.  

The presence of a dip in refractive index profile of a GRIN fiber modifies the mode propagation constants, which disrupts their regular spacing \cite{bib:Tzang:2018sup}, hence the exact periodicity of the collective beam oscillations. Let us consider one of the orthonormal modes of the ideal fiber with parabolic index profile, say $\psi(x,y)$, with a propagation constant $\beta$. Perturbation theory predicts that a dip in the refractive index $\delta n\exp(-r^2/\Gamma^2)$ leads to a shift $\delta\beta$ in the mode propagation constant, which reads as

\begin{equation}
\delta\beta\simeq\frac{\delta n\omega}{c}\int_S|\psi|^2e^{-r^2/\Gamma^2} dS,
\label{eq:pert}
\end{equation}
where $r^2=x^2+y^2$, S is the transverse domain, and $\Gamma$ measures the size of the localized dip. 

According to Eq.\ref{eq:pert}, the presence of a barrier in the potential well (that is, a dip in the refractive index profile) can be interpreted as an astigmatism, which principally affects low-order modes with an intensity peak at the center of the fiber core. On the other hand, the self-imaging period of high-order modes, as well as of any mode with a zero in the core center, will be only weakly affected by the presence of the dip (for more details, please refer to the Appendix). 
Because of their propagation constant shift, low-order modes will exhibit a slightly different beating period than high-order modes. As a result, different modes get out of step,  instead of breathing together in unison. The resulting mode breathing dynamics is graphically visualized in Fig.\ref{fig5}. Here we can see that, in the absence of a dip (panel (a) of Fig.\ref{fig5}), a regularly spaced longitudinal refractive index grating is induced by the Kerr effect. 
On the other hand, in the presence of the refractive index dip (panel (b) of Fig.\ref{fig5}), the longitudinal strictly periodic beam oscillation is disrupted. Only the external crown is regularly spaced, whereas the beam center exhibits three different, and not well localized, domains of index variation.
The disruption of the collective beam oscillations strongly reduces the efficiency of dynamic QPM or GPI as a mechanism for sideband generation. In this case, phase-matching between the pump and sidebands necessarily requires that different spatial modes are involved in the FWM process, as it occurs in IMFWM \cite{bib:MafiJosab:16,bib:Dupiol:17,bib:Conforti:17}.

The phase matching condition for IMFWM can be derived as follows. Supposing that the pump (P) wave is carried by a given mode $p$, the Stokes (S) wave
is carried by a given mode $s$, and that the antiStokes (A) wave is carried by mode $a$, the phase matching condition of IMFWM is $2\beta_p(\omega_{p})-\beta_s(\omega_{s})-\beta_a(\omega_{a})=0$, where $\omega_{a,s}=\omega_p\pm\Omega$. 
By fitting the frequency dependence of the propagation constants as Taylor expansions up to the second order, while assuming the same group velocity
1/$\beta'$ and the same group velocity dispersion $\beta''$ 
(both calculated at pump carrier frequency) for all modes, one obtains 
$\beta_s(\omega_{s})=\beta_s(\omega_{p})-\Omega\beta'+\Omega^2\beta''/2$ and
$\beta_a(\omega_{a})=\beta_a(\omega_{p})+\Omega\beta'+\Omega^2\beta''/2$.
So the phase matching condition can be written as $\Delta\beta=\Omega^2\beta''$, where $\Delta\beta=2\beta_p(\omega_{p})-\beta_s(\omega_{s})-\beta_a(\omega_{a})$
is the difference in propagation constants between the three modes $p$, $s$ and $a$ at the pump carrier frequency, and it grows larger with the distance in modal order between modes.  
The sideband frequency detuning $\Omega$ obeys the condition 
$\Omega^2=\Delta\beta/\beta''$. The higher the frequency detuning, the higher $\Delta\beta$, and consequently the mode order difference.


In the particular case of a GRIN fiber, the mode propagation constants can be associated to  Hermite-Gauss eigenfunctions 
$H_{p,q}(x,y)$. In this case, the frequency detuning can be explicitly written as a function of fiber parameters
$\Omega_h^2=2\sqrt{2\Delta}h/(R\beta'')$, where $R$ is the fiber radius, 
$\Delta$ the relative refractive index difference and  $h$ is an integer measuring the sideband order. 
Incidentally, one can rewrite the same result by replacing $\beta''$ by the refractive index dispersion upon wavelength $n''(\lambda)$. 
In that case, one has $\Omega_h^2/(2\pi)^2=c^2\sqrt{2\Delta}h/(R\lambda^3n'')$ \cite{bib:MafiJosab:16}.
Interestingly, the same values of the sideband shifts $\Omega_h$ also result from GPI, which involves the modulation instability of the longitudinally breathing beam in the ideal GRIN fiber. The crucial difference however, is that GPI sidebands are all carried by the same spatial beam shape as the pump. Whereas in IMFWM pumped by a self-cleaned beam, sideband waves can be generated in higher-order modes.

\begin{figure}[hbt!]
\centering\includegraphics[width=7cm]{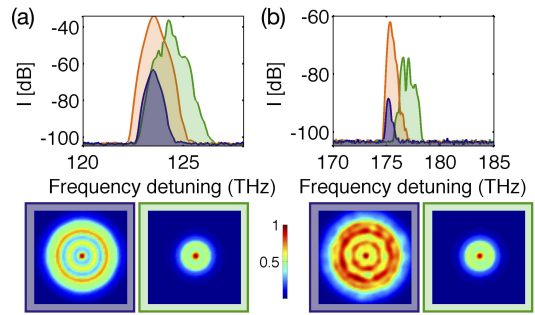}
\caption{Numerical simulation of sideband generation for an input intensity of 10\,$GW/cm^2$, pulse duration of 9\,ps, input beam diameter of 40\,$\mu m$ and propagation distance of 0.16\,m. First (a) and second (b) anti-Stokes sidebands obtained for $\delta n=0$ (green), $\delta n=-0.002$ (orange) and  $\delta n=-0.004$ (violet). The lower frames show the corresponding beam shapes integrated over a 10 THz bandwidth, for the two extreme cases of $\delta n=0$ (green frame) and  $\delta n=-0.004$ (violet frame). \label{fig2}}
\end{figure}

Other forms of perturbations may also affect the index profile of a longitudinally invariant fiber: for example, consider core ellipticity. This situation may equally be studied in the frame of a perturbative approach. In this case, beam propagation exhibits two slightly detuned self-imaging periods (one per axis). Although the introduction of a core ellipticity can be seen as a way of splitting each of the FWM sidebands, the additional astigmatism that is brought by ellipticity causes an 
increase of the beam size at the point of minimum waist, with a consequent reduction of the conversion efficiency. 
Note that other forms of perturbations,  
based instead on the longitudinal periodic modulation of the refractive index, have also been recently proposed to induce sideband splitting of GPI sidebands \cite{bib:Carlos:18}. 

\section{Numerical results}
\label{section_three}
In order to analyze the spatial and spectral properties of parametrically generated waves in GRIN MMFs, we solved the 3D NLSE equation accounting for diffraction, dispersion, parabolic index profile, a localized dip, and the Kerr effect, which reads as

\begin{eqnarray}\nonumber
\frac{\partial E}{\partial z}-i\frac{1}{2k_0}\nabla_{\bot }^2 E+i\frac{\kappa''}{2}\frac{\partial^2 E}{\partial t^2}+ik_0\Delta\frac{r^2}{R^2}E\\
\left. -ik_0\frac{\delta n}{n_0} Ee^{-r^2/\Gamma^2}=i\gamma(1-f_R)|E|^2E\right.
\label{eq:gped}
\end{eqnarray}

Numerical calculations (and analytical phase matching estimates) in a GRIN fiber predict \cite{bib:KrupaPRLGPI,bib:MafiJosab:16}
 the generation of first-order sidebands at a frequency shift of 124.5\,THz (125\,THz) from a laser pump at 1064\,nm, with a fiber of core radius R=26\,$\mu m$,
 core index 1.470, and cladding index of 1.457. 
 
In Fig.\ref{fig2} we compare the numerically computed spectrum, for three representative cases with either $\delta n=0$ (green color, no dip), $\delta n=-0.002$ (orange color), and $\delta n=-0.004$ (violet color). We used the relatively high input intensity of 10\,$GW/cm^2$ and a short propagation distance of 0.16\,m, in order to reduce the computational burden. Panels (a) and (b) of Fig.\ref{fig2} show the first-order and second-order anti-Stokes sidebands, respectively. We may note that for the largest dip with $\delta n=-0.004$, the conversion efficiency drops by about 20dB with respect to the case with no dip. Another interesting feature is that the sideband position is only weakly affected by the presence of the dip, since the sideband appears at 123.5\,THz for $\delta n=-0.004$.
 A similar effect is also observed for the second-order sideband, which moves from a 176.6\,THz detuning (in the absence of a dip) to the frequency shift of 175.2\,THz ($\delta n=-0.004$).
 
Nevertheless, in the presence of a dip, such a small variation in the sideband frequency position (1THz or less) is accompanied by a dramatic variation of the sideband spatial mode content. The frames in the lower row of Fig.\ref{fig2}  compare the sideband beam shape in the absence (green frame) and in the presence (with $\delta n=-0.004$, violet frame) of a dip, respectively. Note that the value of $\delta n=-0.004$ is a dip size which is comparable to the actual value of our perturbed GRIN fiber.
Thus the numerical modeling predicts that, by carving a dip of size comparable to the experimental data at the center of the refractive index profile, one obtains a substantial spatial reshaping of the parametric sidebands, at the expense of significantly lower conversion efficiency. 
Note that a more accurate estimate of the sideband shift, expressed in the wave numbers, can be calculated by using Eq.\ref{eq:pert}, when it is applied to the actual modes of the GRIN fiber with a dip (see Appendix). 

%
\begin{figure}[hbt!]
\centering\includegraphics[width=7cm]{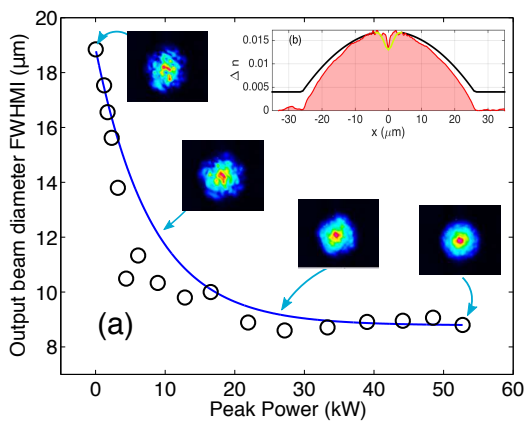}
\caption{(a) Beam diameter vs. input (guided) peak power ($P_{p-p}$)
measured at the full-width at half-maximum intensity (FWHMI), 
emerging from a 10\,m-long GRIN dip MMF. The blue fitting curve is a guide for the eye. 
Insets: output near-field patterns for different $P_{p-p}$. 
(b) Measured index profile of the dip fiber (red filled curve). Parabolic profile used in the simulations (black curve) and gaussian approximation of the dip (yellow curve).} \label{fig1}
\end{figure}

\section{Experimental results}
\label{section_four}
In our experiments, we used two different standard 10m-long GRIN MMFs (core 
diameter 52.1\,$\mu m$ and NA=\,0.205). Their refractive index profiles, illustrated in Fig.\ref{fig:measuredprofiles} of Appendix,
were close to the ideal parabolic shape for the first fiber, and exhibited a dip at the center of the core for the second fiber, as also shown in panel (b) of Fig.\ref{fig1}. We pumped the fibers by an amplified Nd:YAG microchip laser, emitting pulses at 1064\,nm with a temporal duration of 900\,ps and a repetition rate of 30\,kHz. We focused a linearly polarized Gaussian beam at the input face of the fibers with a FWHMI diameter of 30\,$\mu m$. Such coupling method allowed us in principle to excite a large number modes, thanks 
also to the unavoidable presence of random linear mode coupling among all modes. We used two optical spectrum analyzers (OSA) to cover the spectral range from 350\,nm up to 2500\,nm, as well as a CCD camera and beam profiler to study the spectral and spatial reshaping of the multimode beams.

At first, we investigated the possibility to observe Kerr beam self-cleaning in a GRIN MMF with a central dip, in order to compare it with the case of a GRIN MMF without  a dip, as previously reported in \cite{bib:KrupaNatPhot}. Figure \ref{fig1}  illustrates the variation, versus input guided peak power $P_{p-p}$, of the output beam diameter from the fiber with a dip, measured at the full-width at half-maximum intensity (FWHMI). As can be seen, for $P_{p-p}$ above 20\,kW the output beam diameter is nearly halved with respect to its value at low powers. The insets of Fig.\ref{fig1} illustrate the corresponding evolution of the spatial shape of the beam for selected input power values. In spite of the presence of the dip in the refractive index profile of the GRIN fiber, Kerr-induced beam self-cleaning towards a bell-shaped spatial pattern could still be clearly observed. This effect took place before the occurrence of any significant spectral broadening. Although our experiments show that Kerr beam self-cleaning still occurs in the presence of a dip in the refractive index profile, quite remarkably (when considering that the dip is only a minor modification to the index profile of the fiber) the threshold power required to observe the effect is $\sim$ 6 times larger than in the case of a GRIN MMF with an ideal parabolic index profile \cite{bib:KrupaNatPhot}. 

\begin{figure}[hbt!]
\centering\includegraphics[width=7cm]{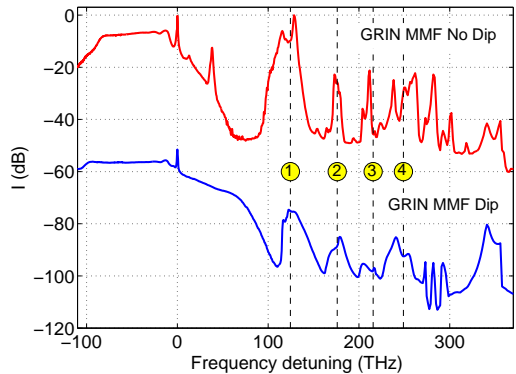}
\caption{Experimental spectra obtained from a 10m-long GRIN MMF without (top, red curve) and with (bottom, blue curve) a central dip in the refractive index profile. The input guided power was $P_{p-p}$=36kW. The vertical dashed lines indicate the analytically calculated sideband frequencies; The blue spectrum was down-shifted of 50dB for better visualization.}
 \label{fig3}
 \end{figure}

Next, we considered the generation of parametric sidebands, which are observed when the input peak power is increased above the threshold for beam self-cleaning. As originally predicted by Longhi \cite{Longhi2003R27} and later experimentally observed by Krupa et al. \cite{bib:KrupaPRLGPI}, the periodic intensity oscillation of the beam due to self-imaging leads, via the Kerr effect, to a dynamic (or light-induced) refractive index grating. This makes the GRIN MMF to behave as a periodic waveguide, which scatters light, via QPM, into unequally spaced sidebands with a large frequency detuning (i.e., $>$100 THz) from the pump, an effect that was named geometric parametric instability, or GPI \cite{bib:KrupaPRLGPI}. In Fig. \ref{fig3} we compare experimental optical spectra obtained in the two multimode GRIN fibers with (blue curve) and without (red curve) the dip, for the input guided peak power of $P_{p-p}$ =36\,kW. 

The vertical dashed lines in Fig.\ref{fig3} indicate the analytically predicted GPI sideband frequencies, calculated from a linear stability analysis of the longitudinally oscillating beam \cite{bib:KrupaPRLGPI}. In GPI, the sidebands are spaced in frequency by $\sqrt{h}$$f_{m}$, where $f_{m}$=124.5THz is the measured frequency detuning for the first resonant sideband, and $h$=1,2,3,4.
As we can see in Fig. \ref{fig3}, with the fiber with a dip in the index profile, the frequency positions of far-detuned anti-Stokes sidebands obey well the values predicted by the theory of GPI \cite{bib:KrupaPRLGPI} or IMFWM \cite{bib:MafiJosab:16,bib:Dupiol:17}.
As already discussed in section \ref{section_two}, the frequencies of the IMFWM 
sidebands coincide with the values of GPI sidebands.
The coincidence of the spectral sideband positions for both fiber samples with the same length and at the same launched power confirms the prediction that the dip has only a minor impact on the sideband positions, in agreement with perturbation theory in section \ref{section_two}, and numerical simulations in section \ref{section_three} (see also the Appendix). 
 
Note however, that although the two fibers had the same diameter, to establish a fair comparison of the
spectral positions of sidebands one would need a precise knowledge of their opto-geometric parameters at the pump wavelength. The dynamics and resolution of our spectrum analyzers was also not sufficient to fully resolve the possible difference in sideband positions.

On the other hand, our experiments demonstrate that the presence of a refractive index dip has a strong impact on the conversion efficiency. The GRIN fiber with a dip led to much lower (reduced by more than 20 dB) frequency conversion efficiencies with respect to the values observed with the ideal GRIN fiber with the same diameter, and for the same launched power. As illustrated in Fig.\ref{fig3}, the intensity difference between the first anti-Stokes sideband and the pump is around 25dB for the GRIN fiber with the dip. This should be compared with the difference of only
3dB between the pump and the first anti-Stokes wave, which is observed with the fiber with a well-defined parabolic index shape \cite{bib:KrupaPRLGPI}.  
The observed large drop in frequency conversion efficiency is well predicted by numerical simulations in section \ref{section_two} (see Fig. \ref{fig2}).

\begin{figure}[hbt!]
\centering\includegraphics[width=7cm]{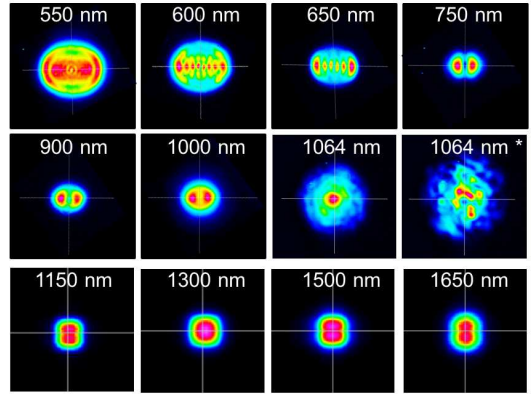}
\caption{Experimental output 2D near-field shapes (normalized intensity in linear scale) versus x (y = 0 section) of a series of selected spectral components measured from 10m-long GRIN MMF with a dip in their index profile at $P_{p-p}$=36\,kW including the first four-orders anti-Stokes parametric sidebands (upper panel); * - near-field shape at the pump wavelength (1064\,nm) in the linear regime with $P_{p-p}$=18\,W. \label{fig4}}
\end{figure}

Most importantly, our experiments clearly demonstrate that only the pump beam at 1064 nm exhibits a spatial distribution with a bell-shaped profile, as a result of Kerr self-cleaning (see Fig. \ref{fig4}). 
A series of transverse beam profiles corresponding to different spectral components, which were selected by using various 10-nm wide bandpass filters with different center wavelengths, are displayed in Fig. \ref{fig4} for the case of the GRIN fiber with a dip in its index profile.
The generated sidebands are carried by high-order modes $H_{q,0}$, with odd values for the index q (see the upper panel of Fig. \ref{fig4}): the larger the order of the peak, the higher the order of the associated spatial mode $q$.
For instance, the beam shape at 750\,nm is close to mode $H_{1,0}$; the beam shape at 650\,nm to mode  $H_{5,0}$, and the beam shape at 600\,nm is close instead to mode $H_{7,0}$.
This observation confirms well the hypothesis formulated in section \ref{section_two}, that phase-matching of the pump and sidebands is achieved via IMFWM. This condition requires the observed increase in mode number as the sideband order grows larger.
Supercontinuum generated by the interplay of stimulated Raman scattering and multimode solitons appearing on the Stokes side of the pump was still observed, however the spatial profile at Stokes frequencies was no longer carried by the fundamental mode, in contrast with the case of a nearly ideal GRIN MMF \cite{bib:SCKrupa}.

\section{Conclusions}
\label{section_five}
In this work, we showed how a small deviation (central dip) from the parabolic profile of the refractive index of a GRIN MMF breaks the dynamic grating periodicity resulting from self-imaging of the pump beam. As a result, the sideband generation mechanism is no longer GPI but IMFWM, and higher-order transverse modes are generated in the sidebands.  
We demonstrated that Kerr self-cleaning of the pump into a bell-shaped beam may still occur, however with a much higher threshold power than with an ideal parabolic profile. Finally, the efficiency of frequency generation is strongly reduced when sidebands are carried by high-order modes. The capability to generate sidebands and supercontinuum light carried by spatial modes of high orders can be of great interest in several applications of imaging and microscopy in the visible and infrared regions, allowing to increase their resolutions. 

\section*{Funding Information}
The European Research Council (ERC) under the European Union's Horizon 2020 research and innovation programme (grant No.~740355); French "Investissements d'Avenir" programme under the ISITE-BFC 299 project (ANR-15-IDEX-0003); ANR Labex ACTION (ANR-11-LABX-0001-01); iXcore research
foundation. K.~K. has received funding from the European Union's Horizon 2020 research and innovation programme under the Marie~Sk\l{}odowska-Curie grant agreement No.~713694 (MULTIPLY). 

\section*{Appendix A: Comparison of the refractive index profiles of the two GRIN fibers}

\begin{figure}[hbt!]
\centering\includegraphics[width=7cm]{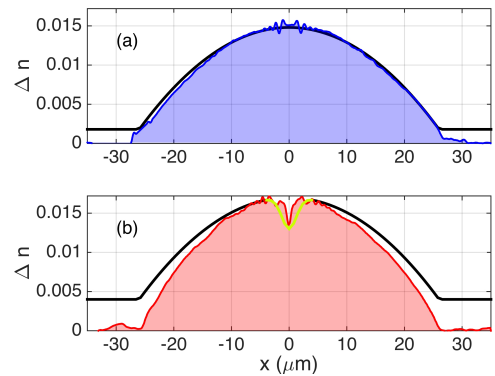}
\caption{Measured refractive index profiles of two types of GRIN fibers.  
(a) fiber with a nearly parabolic profile; the black curve shows the parabolic profile used in the numerical simulations. (b) fiber with a parabolic profile with a dip on the top; the black curves and the yellow curve represent the parabolic profiles and the gaussian approximation of the dip, respectively, as used in the numerical simulations.} \label{fig:measuredprofiles}
\end{figure}

We show in Fig.\ref{fig:measuredprofiles} the refractive index profile of the two GRIN fibers used in the present work.
Panel (a) shows the reference case, where the refractive index is smooth at the top; 
panel (b) instead illustrates the case of the perturbed GRIN fiber, 
where  the presence of a refractive index dip is clearly visible.
 In order to highlight the influence of the dip only, in numerical simulations we approximated the two refractive index profiles by using the same truncated parabola (black curve in Fig.\ref{fig:measuredprofiles}), although the fiber with a dip in panel (b) has a diameter which is slightly smaller
 than the diameter of the  GRIN fiber in panel (a). To simplify the comparison, in the numerical simulations we approximated the dip by a gaussian sinkhole, while keeping for both fibers the same outer diameter (yellow curve in panel (b)).

\section*{Appendix B: Impact of an index dip on the guided modes}

\begin{figure}[hbt!]
\centering\includegraphics[width=7cm]{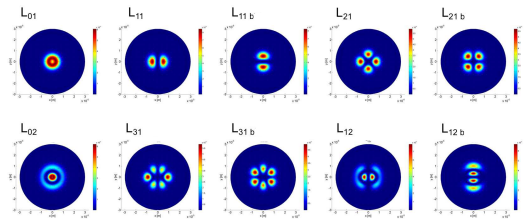}
\caption{Low-order mode profiles for a GRIN fiber with a gaussian dip with $\delta n=-4\times 10^{-3}$. Mode numbering follows the common way of identifying LP modes \label{fig:dipmodes}}
\end{figure}

We show in Fig.\ref{fig:dipmodes} the first 10 guided modes of the GRIN fiber in the presence of a dip, as obtained 
 by a numerical mode solver (COMSOL), by assuming depth of the refractive index dip equal to
  $\delta n=-4\times 10^{-3}$, and a dip radius of 
 1.5\,$\mu m$. In particular, it is possible to observe how the fundamental mode is not a bell-shaped beam, but it exhibits a local intensity reduction, owing to the anti-guiding effect of the dip.

Next, we briefly recall the main steps of the perturbation theory \cite{bib:DMarcuse, bib:Atkins} that we have applied to describe beam propagation in GRIN fibers with a central index dip. 
The complex envelope of the electric field $\tilde{A}(x,y,\omega,z)$, $A(x,y,t,z)$ propagating in a GRIN fiber can be written as
\begin{equation}
2in_0\frac{\omega_0}{c}\frac{\partial\tilde{A}}{\partial z}-\frac{n_0^2\omega_0^2}{c^2}+\nabla_{\bot }^2 \tilde{A}=
-\frac{\omega^2}{c^2}n^2\tilde{A}- \frac{\omega^2}{c^2\epsilon_0}\tilde{f}_{NL}
\label{eq:nlse3d1}
\end{equation}

\noindent The radial ($r$) dependence of the refractive index profile can be approximated in the following way 
\begin{equation}
n(r)=n_0\sqrt{1-2\Delta\frac{r^2}{R^2}}+\delta ne^{-r^2/\Gamma^2}
\end{equation}
where the second term on the right-hand side represents the index perturbation, acting as a barrier for the harmonic potential when $\delta n<0$. In the numerical simulations, the parabolic shape of the core refractive index was truncated once that the core radius was reached.
In Eq.\ref{eq:nlse3d1}, $\tilde{f}_{NL}$ is the term related to the nonlinear response of the fiber.

We may suppose now that the contribution of the Kerr effect in determining the mode profile is of higher-order when compared with the contribution of the waveguide profile. When neglecting nonlinearity, in the CW approximation, and in the absence of a dip, Eq.\ref{eq:nlse3d1} has the following structure 

\begin{equation}
i\frac{\partial A}{\partial z}=-\frac{1}{2k_0}\nabla_{\bot }^2A+k_0\Delta\frac{r^2}{R^2}A
\label{eq:gped2}
\end{equation}
where $k_0=n_0\omega_0/c$.
Equation \ref{eq:gped2}  can be also written as:
\begin{equation}
i\frac{\partial A}{\partial z}=-\frac{1}{2k_0}\nabla_{\bot }^2 A+V_0(r)A={\cal H}^{(0)}A
\label{eq:nlse3d}
\end{equation}

\noindent $V_0(r)=k_0\Delta r^2/R^2$ is the harmonic potential (in the approximation of a parabolic profile of the refractive index), $r^2=x^2+y^2$, and ${\cal H}^{(0)}$ is the Hamiltonian operator of the unperturbed waveguide (in numerical simulations we assumed that the potential is a truncated paraboloid).

We may model the presence of a dip by adding to the harmonic potential a small gaussian barrier $V(r)=V_0(r)+\epsilon\exp(-r^2/\Gamma^2)$. In this case, $\epsilon=-k_0\delta n/n_0$ is a measure of the depth of the dip and $\Gamma$ fixes the dip radius.

The new equation reads as
\begin{equation}
i\frac{\partial A}{\partial z}={\cal H}A=({\cal H}^{(0)}+{\cal H}_1)A
\label{eq:perturbed}
\end{equation}
where ${\cal H}_1=\epsilon\exp(-r^2/\Gamma^2)$ is the perturbation term of the dip. Eq.\ref{eq:perturbed} looks formally equivalent to the previous one, under the influence of a static perturbing force: this is the common procedure to study the influence of a perturbation such as an electric or magnetic field, that can change the energy levels and the shape of orbitals (the propagation constants and the shape of the guided modes in our case).
The perturbation theory can thus be of help in identifying the consequences brought by the dip in the propagation constants, and in the shape of the guided modes \cite{bib:DMarcuse, bib:Atkins}.

Let us suppose now that the guided modes of the {\it perturbed} system ${\cal H}$ are of the form $A=\psi(x,y)\exp(-i\beta z)$, where $\beta$ is the mode wave-vector shift from the reference wave-vector of the core. A first insight on how the propagation constants are modified can be obtained by the Hellmann-Feynman theorem which says, once applied to the depth $\epsilon$ of the gaussian barrier, that
\begin{equation}
\frac{\partial\beta}{\partial\epsilon}=
\int\psi^*\left(\frac{\partial {\cal H}_1}{\partial\epsilon}\right)\psi dxdy=\int\psi^*(e^{-r^2/\Gamma^2})\psi dxdy
\label{eq:HF}
\end{equation}
 Although the theorem requires the modes $\psi$ of the {\it perturbed} problem, it is already evident the key point that the largest consequences for the propagation constants are for those modes with a field maximum around the dip.  Modes with nearly zero intensity in the center will be much less affected. Note that the degeneracy and the equal spacing among wave-vectors are a basic ingredient of GPI \cite{bib:KrupaPRLGPI} in GRIN fibers. Therefore, since the presence of an index defect directly perturbs the equal spacing of wave-vectors, as well as the overlap integrals, the selection rules of the parametric processes will also be perturbed.  

As a confirmation, we obtain from the numerical mode solver the difference in propagation constants for the fundamental mode in absence and in presence of dip as $\beta_{H00,ideal}-\beta_{LP01,dip}=1.98\times 10^{3}m^{-1}$. The difference in propagation constants for the first odd-parity mode in the presence and in the absence of dip is reduced by nearly of one order of magnitude: $\beta_{H10,ideal}-\beta_{LP11,dip}=3.83\times 10^{2}m^{-1}$.

The dip-induced perturbation has the direct effect of modifying the separation between adjacent group of modes. However, the corresponding analysis would require using a second-order perturbation theory, since the frequency detuning depends upon the difference between pairs of propagation constants. In fact, depending on the relative depth of the local defect, one can modify not only the propagation constants but also the shape of the modes (e.g. the fundamental mode presents a local depression in the center). Thus one obtains that the synchronized 
beating of modes is gradually spoiled, with a reduction of the collective effect or self-imaging of the beam as described in Ref. \cite{bib:KrupaPRLGPI}.

\begin{figure}[hbt!]
\centering\includegraphics[width=7cm]{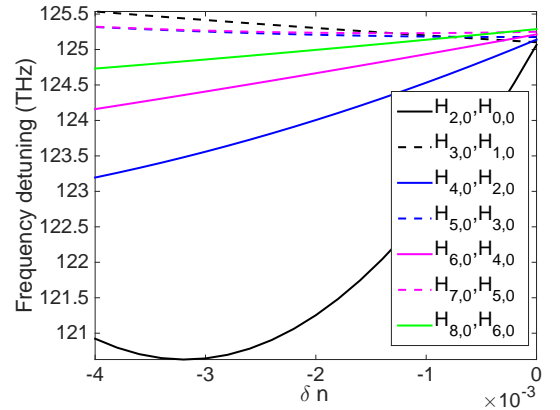}
\caption{Calculated shift of the sideband positions upon barrier depth. \label{fig:pertfreqdet}}
\end{figure}

Perturbation theory permits to estimate the propagation constants and the shape of modes in the presence of the gaussian barrier, or dip,
starting from the solutions of the ideally parabolic GRIN fiber. The procedure begins by solving the mode equation for the unperturbed GRIN fibers with Hamiltonian ${\cal H}^{(0)}$, which in Dirac notation for the unperturbed wave-vector $\beta_{0,n}$ of mode $\psi_{0,n}$, reads as
\begin{equation}
{\cal H}^{(0)}|n>=\beta_{0,n}|n>
\end{equation}
Now we consider ${\cal H}={\cal H}^{(0)}+\mu {\cal H}^{(1)}$, where $\mu$ is a book-keeping parameter to help separating the different orders.
We consider also $\psi_{n}=\psi_{0,n}+\mu \psi^{(1)}_{0,n}$: this expression shows how the guided mode $\psi_{n}$
can be obtained by adding to the unperturbed mode solution a correction term at first order $\psi^{(1)}_{0,n}$.
Similarly $\beta_{n}=\beta_{0,n} +\mu\beta^{(1)}_{0,n}$.
By applying this formalism to the equation ${\cal H}\psi_{n}=\beta_{n}\psi_{n}$, and after separating terms of the same order in $\mu$, we obtain at order zero
\begin{equation}
{\cal H}^{(0)}\psi_{0,n}=\beta_{0,n}\psi_{0,n};
\end{equation}
whereas at first order one has
\begin{equation}
({\cal H}^{(0)}-\beta_{0,n})\psi^{(1)}_{0,n}=(\beta^{(1)}_{0,n}-{\cal H}^{(1)})\psi_{0,n}.
\end{equation}
It is customary to develop the perturbation as a linear combination of the eigenmodes of the unperturbed problem, which is our case are the guided modes of the unperturbed GRIN fiber
$\psi^{(1)}_{0,n}=\sum_h a_h\psi_{0,h}$.
After some algebra, it is possible to derive the known result of first order perturbation theory
\cite{bib:Atkins}:
\begin{equation}
\beta^{(1)}_{0,n}=\int\psi^*_{0,n}({\cal H}^{(1)})\psi_{0,n}dxdy.
\end{equation}
With the same procedure, it is also possible to calculate the correction in the modal shape induced by the index dip, which reads as
\begin{equation}
\psi^{(1)}_{n}=\psi_{0,n}+\sum_h\frac{<h|{\cal H}^{(1)}|n>}{\beta_{0,n}-\beta_{0,h}}\psi_{0,h}
\label{eq:1storderpsi}
\end{equation}
The numerators in the sum appearing in eq.\ref{eq:1storderpsi} account for the overlap among modes and 
the physical deformation of the refractive index: the influence of a central dip is expected larger for those modes having a maximum at the center of the fiber, as already anticipated by calculating Eq.\ref{eq:HF}. 
Moreover, looking at the denominator, the terms tend to vanish as the difference of the wave-vectors grows larger. 
The above procedure can be extended to the case of degenerate modes \cite{bib:Atkins,bib:DMarcuse}.

In GPI, the key ingredient is the beating among modes of adjacent group orders. As another outcome of the perturbation theory, we may calculate the beating among two modes of the same parity, here emphasized by the choice of mode indexes n and n+2:
\begin{equation}
\begin{split}
\beta_{n+2}-\beta_{n}=\beta_{0,n+2}-\beta_{0,n}\\
+ \left\{  \int\psi^*_{0,n+2}({\cal H}^{(1)})\psi_{0,n+2}dxdy-\int\psi^*_{0,n}({\cal H}^{(1)})\psi_{0,n}dxdy \right\}
\label{eq:correctospacing}
\end{split}
\end{equation}

Although eq.\ref{eq:correctospacing} was calculated from a simple first order perturbation theory, it shows the  intuitive result that the refractive index dip may perturb each wavenumber in a different way. This fact can in turn affect  the equal spacing of wavenumbers with consequences in the parametric generation of sidebands. The term in brackets in Eq.\ref{eq:correctospacing} vanishes in the absence of perturbation or for a perturbation equally affecting the modes. Whereas the remaining term $\beta_{0,n+2}-\beta_{0,n}$ is responsible for the self-imaging properties of ideal GRIN fibers. Cautions should be taken because eq.\ref{eq:correctospacing} represents the difference between two wavevectors, each of them calculated with the perturbative method. Should the error be relevant one can proceed by extending the perturbative analysis to the second order, so to have a more reliable estimation of the propagation constants in presence of perturbation \cite{bib:Atkins}. The effect of the perturbation term can be considered as a sort of astigmatism (or better aberration), since pairs of modes have different beat lengths, in analogy with free-space optical systems, where rays of perpendicular planes can have different foci.   Note also that high-order modes are only weakly affected by the perturbation, and collective effects remain possible among them. Low-order modes (among which the fundamental mode) instead are highly affected by the term in brackets, because the influence of the dip varies a lot among these modes. We expect that low-order modes will have then a slightly different beating period than that of high-order modes, so that they get de-synchronized instead of breathing together as a whole. 

\begin{figure}[hbt!]
\centering\includegraphics[width=7cm]{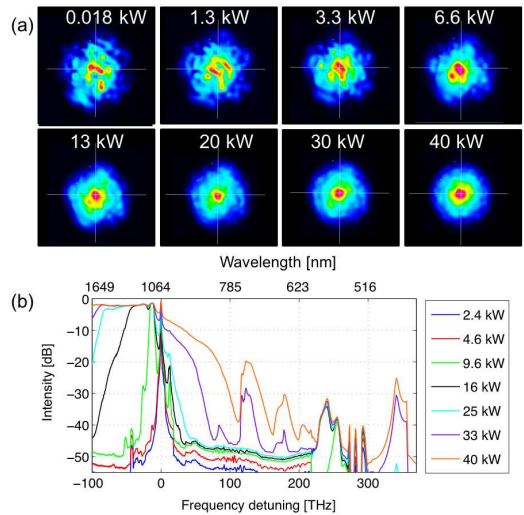}
\caption{Experimental evolutions of 2D near-field shape at 1064 nm (normalized intensity in linear scale) (a) and spectrum (b) as a function of input guided power $P_{p-p}$  measured at the output of 10m-long GRIN MMF with perturbed index profile.
\label{fig:spacespectr}}
\end{figure}

The overall spatial beam dynamics can be visualized (see Fig.\ref{fig5} of the main text) 
by collecting the iso-intensity surfaces, which identify the local variation of the refractive index induced by the Kerr effect. In particular, in the absence of a dip (green diagram), the iso-intensity surfaces show the fingerprint of the longitudinal grating, which is regularly spaced.
In the presence of a dip instead (violet diagram), the profile is broken by the aberration brought by the dip: only the external crown is regularly spaced, whereas the center presents three different and not well localized zones of intensity variation.

The perturbative analysis permits to explore the impact of any given parameter. For instance, one can 
repeat the analysis by obtaining an approximate estimation of the frequency shift for each pairs of beating modes upon the barrier depth $\delta n$ . 
Results of these approximated computations (at the second order)
are given in Fig.\ref{fig:pertfreqdet}. Note that all of these frequencies converge to the same value 
for $\delta n=0$, whereas all frequency modes tends to spread as soon as the barrier depth increases in modulus.
From these results, it is reasonable to expect that parametric sidebands will broaden, since modes taken by pairs may generate the parametric frequencies in slightly different positions.

\section*{Appendix C: Additional experimental results}

Figure \ref{fig:spacespectr} illustrates the evolution of the spatial shape of the pump beam (panel (a)) measured at the output of the GRIN MMF with a dip, 
as a function of input guided peak power $P_{p-p}$. Panel (b)
 of the same figure illustrates how the overall spectrum (shown upon wavelength and upon frequency detuning from the pump) varies with the input guided power.

\begin{figure}[hbt!]
\centering\includegraphics[width=7cm]{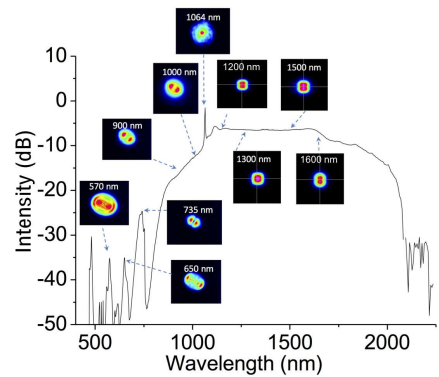}
\caption{Supercontinuum generation in the GRIN fiber with dip: fiber length of 10\,m and average power of 955mW. We used a series of 10-nm wide bandpass filters with center wavelengths of 550nm, 650nm, 750nm, 900nm, 1000nm, 1200nm, 1300nm, 1500nm, 1600nm, and 3-nm wide bandpass filter at 1064nm.\label{fig:dip_figre_SC_Vincentmeasures_10m_955mW}}
\end{figure}

Figure \ref{fig:dip_figre_SC_Vincentmeasures_10m_955mW} summarizes the supercontinuum spectrum and the associated transverse beam shapes at some selected wavelengths from the GRIN fiber with dip. In this different series of 
experiments it is visible how the pump beam only has a bell-shaped beam, whereas the supercontinuum is carried mainly by a beam shape with two lobes. This result is of particular interest, and it shows how a defect in a graded index fiber can lead to the generation of a supercontinuum with a beam shape with two main lobes.
 The change in orientation of the lobes for wavelengths above 1064\,nm simply 
indicates the use of another camera (InGa	As).  The sharp parametric sidebands at wavelengths lower than the pump 
show more complex profiles, with additional lobes as the sideband wavelength decreases, as discussed in the main text of the paper.

\begin{figure}[hbt!]
\centering\includegraphics[width=7cm]{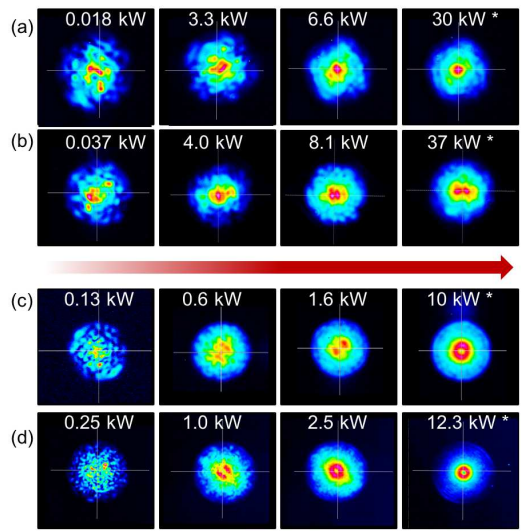}
\caption{Experimental output 2D near-field shapes (normalized intensity in linear scale) as a function of input guided power $P_{p-p}$ measured at the pump wavelength of 1064 nm ((a)-(c)) and 532 nm (d) in 10m-long ((a), (b)) and 30m-long ((c), (d)) GRIN MMF. * - results for the $P_{p-p}$ at which frequency conversion into sidebands was also observed.
 \label{fig:allclean}}
\end{figure}

Figure \ref{fig:allclean} illustrates the variation of the spatial shape measured at the output of the GRIN MMF with a dip, as a function of input guided peak power $P_{p-p}$ at the fundamental wavelength. 
In panel (a) of Fig.\ref{fig:allclean}, we may clearly observe the Kerr-induced beam self-cleaning effect \cite{bib:KrupaNatPhot} towards a bell-shaped spatial beam profile. This spatial reshaping takes place before the occurrence of any significant spectral broadening. When slightly adjusting the input excitation conditions, we could observe that an initially multimode beam, when increasing its input power, could self-organize towards a spatial shape which bears strong resemblance to mode $H_{0,1}$, as it is shown by the frames of panel (b) in Fig.\ref{fig:allclean}. 
Note that similar odd-parity beam shapes were also experimentally observed in a more systematic study carried out with the ideal GRIN fibers in Ref.\cite{bib:EtienneArxiv}, and also numerically predicted in Ref. \cite{Kivshar2018multimode}, in a different context of multimode solitons.

In another series of experiments, we used shorter pulses (30\,ps duration), which were delivered by a flash-pumped Nd-YAG laser emitting at 1064nm or at its second harmonic of 532 nm, with a repetition rate of 20 Hz. In that series of experiments, we considered spans of the dip perturbed GRIN MMF with a length of 30 m. The experimental results of spatial reshaping of the fundamental beam are displayed in panels (c) and (d) of Fig.\ref{fig:allclean} for the pump wavelengths of 1064\,nm and 532\,nm, respectively. 
In the linear regime, the speckle sizes are related to the guided modes, whose number increases by reducing the carrier wavelength. As a consequence, for the pump at 532\,nm the speckle grains are far smaller than those observed at 1064\,nm.

Nevertheless, also for the second harmonic pump at 532\,nm we observed the effect of spatial self-cleaning. 
As presented in Fig.\ref{fig:allclean}(c) and (d), similarly to the results previously discussed with 900\,ps pulses,
 low-power speckled beams at both fundamental waves initially evolve into high-power  modes close to $LP_{11}$, but then end up by forming a doughnut shape, which is further transferred across all spectral range generated via Raman scattering when the input power grows larger. Such doughnut shape is likely related to the fundamental mode of the dip fiber, as it is calculated by the mode solver. Our experimental results clearly show that by modifying the input conditions it is also possible to isolate a doughnut shape, which can be of interest for imaging applications.

\bibliography{BIBLIO_dip_combo.bib}

\end{document}